\documentclass[aps,pre,twocolumn,superscriptaddress]{revtex4-1}
\usepackage{epsfig}
\usepackage{amsmath}

\begin{document}

\title{Heterogeneous update mechanisms in evolutionary games: mixing innovative and imitative dynamics}

\author{Marco Antonio Amaral}
\affiliation{Departamento de F\'\i sica, Universidade Federal do Rio Grande do Sul - RS, Brasil}

\author{Marco Alberto Javarone}
\affiliation{School of Computing, University of Kent, Chatham Maritime, UK}
\affiliation{nChain Ltd, London W1W 8AP, UK}
\affiliation{School of Computer Science, University of Hertfordshire, Hatfield AL10 9AB, UK}

\begin{abstract}
Innovation and evolution are two processes of paramount relevance for social and biological systems. In general, the former allows the introduction of elements of novelty, while the latter is responsible for the motion of a system in its phase space. Often, these processes are strongly related, since an innovation can trigger the evolution, and the latter can provide the optimal conditions for the emergence of innovations. Both processes can be studied by using the framework of Evolutionary Game Theory, where evolution constitutes an intrinsic mechanism. At the same time, the concept of innovation requires an opportune mathematical representation. Notably, innovation can be modeled as a strategy, or can constitute the underlying mechanism which allows agents to change strategy. Here, we analyze the second case, investigating the behavior of a heterogeneous population, composed of imitative and innovative agents. Imitative agents change strategy only by imitating that of their neighbors, whereas innovative ones change strategy without the need of a copying source. The proposed model is analyzed by means of analytical calculations and numerical simulations in different topologies. Remarkably, results indicate that the mixing of mechanisms can be detrimental to cooperation near phase transitions. In those regions, the spatial reciprocity from imitative mechanisms is destroyed by innovative agents, leading to the downfall of cooperation. Our investigation sheds some light on the complex dynamics emerging from the heterogeneity of strategy revision methods, highlighting the role of innovation in evolutionary games.
\end{abstract}

\maketitle

\section{Introduction}
\label{Introduction}
The emergence of cooperation is a topic of paramount relevance in different areas, as demonstrated by the long list of contributions across various fields, ranging from biology to sociology, and from economics to robotics  \cite{Pennisi2005,Nowak2011a,Perc2017,Hauert2005}. In a broad sense, why should people cooperate with their peers in a competitive scenario, where selfish individuals would often fare better? Evolutionary Game Theory (hereinafter EGT) constitutes one of the most suitable tools for approaching such question \cite{Smith82, Axelrod1984,Weibull1995}, and the Prisoner's Dilemma represents the canonical way for studying how a cooperative behavior can emerge in a competitive scenario \cite{Nowak2006, Perc2017}. The dynamics of evolutionary games show how cooperation results from a collective behavior. Notably, these models consider a population that, under particular conditions, is able to reach an equilibrium of cooperation even when the agent interactions are based on games whose Nash equilibrium is defection.

One of the earliest approaches in EGT, proposed by Maynard Smith~\cite{Smith82, sigmund_tpb05}, uses the mathematical framework of birth-death dynamics usually seen in biological evolution, in a model where individuals copy the strategy of more successful contacts (akin to a Moran process). Using a linear copy probability, this mechanism leads to the classical replicator equation~\cite{Szabo2007}, i.e. the general mathematical model for natural evolution. However, from a Game Theory perspective, individuals can change strategy by many other mechanisms, e.g. imitation of the best, win-stay-lose-learn, tit-for-tat, and so on and so forth~\cite{Perc2017, Szabo2007}. 
Here, updating rules based on imitative mechanisms can be defined as non-innovative~\cite{Szabo2007}, since they allow individuals to choose only among strategies adopted in their neighborhood. As result, once a strategy disappears, it can be considered as extinct if there is no external mutation mechanism. It is important to note how mutation mechanisms can lead to diversity, but they are not directly related to an innovative updating rule, which represents the ability of one individual to choose a strategy that does not appear in its neighborhood.
On the other hand, mechanisms that lead individuals to change strategy without the need to copy from a source (e.g. a neighbor) can be defined as innovative.
For instance, one individual might change strategy by analyzing the trend of her/his gain, e.g. a decreasing gain might lead to test a different strategy.
One of the most famous case is the win-stay-lose-shift, where if the individual has a payoff below some aspiration level, she/he simply changes strategy, no matter which strategies are available from the neighborhood. Two other famous examples of innovative updating rules are the Logit rule and best response~\cite{blume_l_geb95, Amaral2017, szabo_jtb12b, roca_epjb09}.

From the point of view of Information Theory, there is an important difference between these two classes of updating rules, i.e. innovative and non-innovative. Notably, rational individuals~\cite{Javarone2016d} select their strategy according to rules that take into account their gain over time, or their current gain and that of their neighbors. Therefore, they need some information in order to take a decision. Thus, the essential difference between innovative and non-innovative individuals lays in the information source they adopt/have about the system. 
As reported in previous investigations (e.g.~\cite{Amaral2016b, Amaral2017, Armano2017, Xu2017, roca_epjb09}), the application of innovative and non-innovative updating rules leads to results that can be drastically different. As a very interesting and recent example, \cite{Kazuki2018} showed how innovative strategies towards vaccination can lead to different dynamics than the usual imitative ones, changing the vaccination coverage.
It is interesting to observe that imitative mechanisms are usually associated with long term biological evolution. The copying process is related to the offspring of a more successful individual inheriting her/his parent strategy. On the other hand, innovative dynamics can model the behavior of individuals that, like humans, have cognitive responses to the environment~\cite{vukov_njp12, bonawitz_cg14}, and that happen on shorter time scales.

Imitative dynamics have been broadly studied for different games, updating rules and connection topologies. Some classic mechanisms that support cooperation in this setting include kin selection \cite{Hamilton1964}, mobility and dilution \cite{sicardi_jtb09}, direct and indirect reciprocity \cite{axelrod_s81}, network reciprocity \cite{Nowak1992a, santos_pnas06}, group selection \cite{wilson_ds_an77}, dissociation \cite{wardil_jpa11}, and population heterogeneity \cite{szolnoki_epl07, perc_pre08, santos_jtb12} (for reviews see \cite{Szabo2007, perc_bs10, perc_jrsi13}). Nevertheless, innovative mechanisms still require deeper studies in the evolutionary context.
It is worth highlighting the strong relation between innovation and cooperation, as reported in recent works demonstrating that, if a population adopts just an innovative strategy for performing updates, cooperation can be sustained for a large range of parameters. In~\cite{Amaral2016b}, the authors show that win-stay-lose-shift with dynamic aspiration can lead to the coexistence of cooperation and defection for the whole parameter range while, at the same time, cooperators do not need to form islands to survive. In~\cite{Amaral2017, szabo_jtb12,Danku2018}, a model based on the Glauber dynamics (from magnetism) shows the survival of cooperators while leading the population to global stable patterns, in a process akin to the minimization of energy.
Driven by this observation, in this work we propose an evolutionary model for studying the dynamics of a heterogeneous population composed of imitative and innovative agents. In particular, imitative agents adopt the typical copying mechanism with a probability weighted by the Fermi-Dirac distribution, while the innovative ones use the Logit rule (also weighted by the Fermi-Dirac distribution).

Heterogeneity, in the most general form, is a strong facilitator of cooperation. The mixing of strategies, different kinds of players, topologies, etc, has been shown time and again to be a great promoter of cooperation~\cite{Kokubo2015, santos_jtb12, Amaral2015, Amaral2016, Qin2017, zhu_p_pone14, tanimoto_pre07b, perc_pre08, szolnoki_epjb08, iwa_pha15, wardil_csf13, liu_rr_epl15}. In this sense, we mix two kinds of agents, each one following a specific updating rule. In doing so, we can analyze the results coming from a form of heterogeneity related to the ``updating rules''.

We first solve the mean-field equation for the model in the well-mixed case and perform Monte-Carlo simulations in a square lattice to observe the effects of the spatial structure. Most intriguingly, we find that while a pure innovative population can maintain a high level of cooperation, a minimum cooperation level occurs in the mixed state of innovators and imitators. This happens for the Prisoner's Dilemma near a critical point that characterizes the phase transition of the pure imitative model. In order to verify the robustness of this result, we also analyze other connection topologies and games like the Stag-Hunt and Snow-Drift. Lastly, we study what mechanisms create this drop in cooperation for the mixed states using lattice snapshots and the individual fraction of each population (i.e. innovative cooperators, innovative defectors, etc). 

The remainder of the paper is organized as follows: Section~\ref{Model} introduces the proposed model and its dynamics; Section~\ref{results} reports results of analytical calculations and numerical simulations. Finally, Section~\ref{conclusion} provides a summary of the main outcomes and related observations.

\section{Model}
\label{Model}
In the proposed model, we aim to clarify the influence of innovation in the dynamics of evolutionary games, by considering the behavior of a population whose agents have two strategies available, cooperation ($C$) and defection ($D$). Such scenario can be represented by the following payoff matrix:
\begin{equation}\label{paymatrix}
 \bordermatrix{~ & C & D \cr
                  C & R & S \cr
                  D & T & P \cr},
\end{equation}
\noindent where two cooperative agents receive a reward ($R$), two defectors receive a punishment ($P$), and an agent that cooperates with a defector receives $S$, while the defecting agent receives a temptation ($T$). Using the parametrization $R=1,P=0,S=[-1,1],T=[0,2]$, we can explore the dynamics of the model in four different configurations, i.e. the Prisoner's Dilemma for ($T>1,S<0$), the Stag-Hunt for ($T<1,S<0$), the Snow-Drift for ($T>1,S>0$), and the Harmony Game for ($T<1,S>0$)~\cite{Szabo2007, roca_pre09}. 

In addition, our agents are provided with a character, i.e. they can be innovators or imitators. Notably, a fixed fraction of agents, say $\alpha$, will update its strategy according to a mechanism based on innovation. Whereas, a fraction $(1-\alpha)$ of agents will change its strategy by adopting the typical imitative dynamic~\cite{Nowak2006, Szabo2007}. We emphasize that while agents can change strategy (e.g. from $C$ to $D$) over time, their character (imitative or innovative) never changes.
As result, an imitative agent $i$, at each update, randomly chooses one neighbor $j$ and copies its strategy with probability:
\begin{equation}\label{imit}
		p(\Delta u_{ij})_{imt}=\frac{1}{ 1+e^{-(u_j-u_{i})/k} }~,
\end{equation}
\noindent where $u_i$ and $u_j$ indicate the payoff of the selected agent ($i$) and of its neighbor ($j$), respectively, while $k$ represents the agent's irrationality. 
Here we set $k=0.1$, i.e. a numerical value that is widely used in the literature in order to add small noise in the decision process. As above reported, the imitation rule is a non-innovative mechanism \cite{Szabo2007}, because an agent changes strategy by considering only among those available in its neighborhood. In doing so, new strategies can never appear once extinguished and, most importantly, agents can never ``explore'' new ones~\cite{roca_epjb09, Szabo2007}. Notably, the process of imitation is similar to local competition where death is a random uniform process, and reproduction rates are determined by the payoff (fitness).

On the other hand, an innovative agent, $i$, changes its current strategy to the opposite one with probability:
		\begin{equation}\label{logit}
		p(\Delta u_{i})_{inv}=\frac{1}{ 1+e^{-(u_{i*}-u_{i})/k} }~,
		\end{equation}
\noindent where $u_{i*}$ is the agent's own payoff if it had changed to the opposite strategy and everything else remained the same. It is worth to remind that this updating rule corresponds to the Glauber dynamics in magnetic models \cite{Glauber1963, binder_prb80} while, in the context of Game Theory, it is known as Logit dynamics, myopic best response or myopic Logit rule \cite{blume_l_geb95, Danku2018, roca_epjb09}. According to this rule, an innovative agent evaluates the gain that it might achieve by changing strategy, under the hypothesis that its neighborhood remains unchanged. 
As reported in previous investigations~\cite{Amaral2016b, Amaral2017, szabo_jtb12b, roca_epjb09, Sysi-Aho2005} this mechanism leads to very different results compared to imitative dynamics. 
Tuning the value of $\alpha$ between $0$ (i.e. full imitation) and $1$ (i.e. full innovation), we aim to analyze how innovation affects the dynamics towards cooperation, in different conditions.

\section{Results}\label{results}
The proposed model is studied by means of numerical simulations, by arranging agents over a regular lattice and on complex networks. However, as a preliminary study, we perform analytical calculations considering the dynamics of a well-mixed population using the mean-field approximation.
\subsection{Well-mixed population}
We begin the analysis of the proposed model with the case of a well-mixed population. Notably, by using the master equation in the mean-field approach \cite{Szabo2007, matsuda_h_ptp92,Javarone2016b}, the temporal evolution of the cooperator's density, $\rho$, reads
\begin{equation}\label{master}
\dot{\rho}= (1-\rho) \Gamma_+ - \rho \Gamma _-,
\end{equation}
\noindent where $\Gamma_+$ stands for the average rate at which agents change strategy from $D$ to $C$, leading to an increase in $\rho$ (and similarly to $\Gamma_-$). While usually this rate depends on just one updating rule, in our case we need to consider the presence of two kinds of agents, i.e. innovators and imitators. As result, the both rates, $\Gamma_\pm$, will be the average rate between each updating rule, weighted by $\alpha$:
\begin{equation}
\Gamma_\pm=(1-\alpha)\Gamma_{\pm imt}+\alpha\Gamma_{\pm inv}
\end{equation}

\begin{figure}
\includegraphics[width=7.5cm]{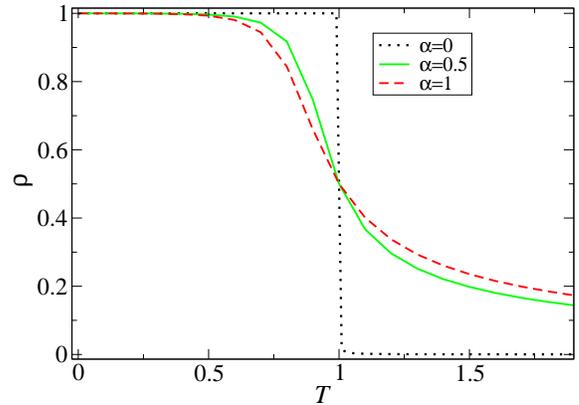}
\caption{Asymptotic cooperation fraction ($\rho$) versus temptation to defect ($T$) in the weak Prisoner's Dilemma ($S=0$) in the mean-field, well-mixed population for the innovative ($\alpha=1$) and imitative ($\alpha=0$) model, compared with the mixed population of half innovative and half imitative agents ($\alpha=0.5$).}
\label{fig_meanfieldeq}
\end{figure}

Notably, for the well-mixed population we have the following rates \cite{Szabo2007, Amaral2017},
\begin{eqnarray}\label{gamma}
\Gamma_{+imt}= \frac{\rho}{1+e^{-A/k}}~,\\
\Gamma_{-imt}= \frac{1-\rho}{1+e^{+A/k}}~,\\
\Gamma_{\pm inv}= \frac{1}{1+e^{\mp A/k}}~,
\end{eqnarray}
\noindent where $A=\rho(1-T)+(1-\rho)S$ is the difference in the average payoff from a typical $C$ and $D$ agent interacting with all other agents. Note that for the innovative updating rate, $\Gamma_{\pm inv}$, the only change between the positive ($\Gamma_{+ inv}$) and negative ($\Gamma_{- inv}$) rate is in the sign of $A$. Accordingly, the full equation becomes
\begin{equation}\label{master2}
\dot{\rho}= (1-\rho)\left[  \frac{\rho(1-\alpha)+\alpha}{1+e^{-A/k}}\right]  - \rho\left[\frac{(1-\rho)(1-\alpha)+\alpha}{1+e^{A/k}}\right] .
\end{equation}
We solve Eq.~(\ref{master2}) numerically, letting the system reach the equilibrium point as $t\rightarrow \infty$. This gives the asymptotic  behavior of the population for the well-mixed case ---see Fig.~\ref{fig_meanfieldeq} for the fully imitative ($\alpha=0$), fully innovative ($\alpha=1$) and mixed ($\alpha=0.5$) cases.

This preliminary analysis shows that the behavior of the heterogeneous population is not just the average value of the two pure cases ($\alpha=0$ or $\alpha=1$), i.e. even if half population is imitative, the behavior is much more similar to the pure innovative population. Also, we observe in Fig.~\ref{fig_meanfieldeq} that the point $T=1$ is relevant, as it defines which updating rule leads to the highest value of cooperation. If $T<1$, the imitative population has a higher cooperation fraction, while for $T>1$, the innovative population has higher levels of cooperation. Specifically, if $T=1$ and $S=0$, we obtain that $A=0$ in Eq. \ref{master2}, leading to $\dot{\rho}=\alpha(1-2\rho)$. This ODE has only one fixed point at $\rho=0.5$, that is independent of $\alpha$, as we see in Fig.~\ref{fig_meanfieldeq}; all three models have the same value of $\rho$ for $T=1$. 

\begin{figure}
\includegraphics[width=7.5cm]{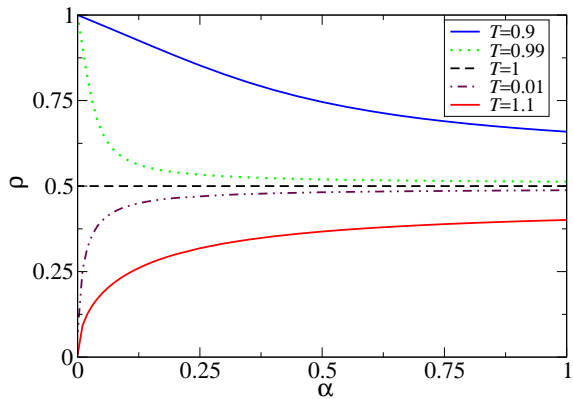}	
\caption{Asymptotic cooperation level, $\rho$, versus fraction of innovators, $\alpha$, here $S=0$. Results obtained from the mean-field equation in the well-mixed population. The increasing of innovators is beneficial for cooperation in the weak Prisoner's Dilemma ($T>1$), while detrimental to it in the region $T<1$. In $T=1$ all models have the same asymptotic behavior.}
\label{fig_alphaedo}
\end{figure}

We proceed analyzing how different values of $\alpha$ affect the population. In particular, as shown in Fig.~\ref{fig_alphaedo}, cooperation increases monotonously with $\alpha$ in the region $T>1$, while the opposite occurs for $T<1$. In this case, innovation is beneficial to cooperation only for the Prisoner's Dilemma region of the parameter  $T$. If $T<1$, which characterizes the region of Stag-Hunt and Harmony-Game, cooperation fares better if there are more imitative agents, i.e. low $\alpha$ values.

\subsection{Structured population}
In order to study the behavior of the proposed model considering a structured population, we initially perform Monte Carlo simulations by arranging $10^4$ agents in a square lattice with periodic boundary conditions.
Here, at each time step, an agent (say $i$) interacts with its neighbors and, according to the payoff matrix of the game, obtains a cumulative payoff. Then, agent $i$ undergoes the `strategy revision phase' (SRP) that is based on the probability defined in Eq.~(\ref{imit}), or in Eq.~(\ref{logit}), depending on its nature, i.e. imitator or innovator.  
Thus, the described set of actions (i.e. from the agent selection to the SRP) is repeated $N$ times (where $N$ is the total number of agents), which constitute a single Monte Carlo Step (MCS). The simulation lasts until the population reaches a stable state ($10^3-10^4$ MCS's)~\cite{Szabo2007}. After that, results are averaged over the last $1000$ MCS, and observed for $10-50$ different initial conditions. It is worth reporting that, at the beginning of each simulation, we start with a homogeneous strategy distribution, so that half population is composed of cooperators, and half of defectors.

\begin{figure}
\includegraphics[width=7.5cm]{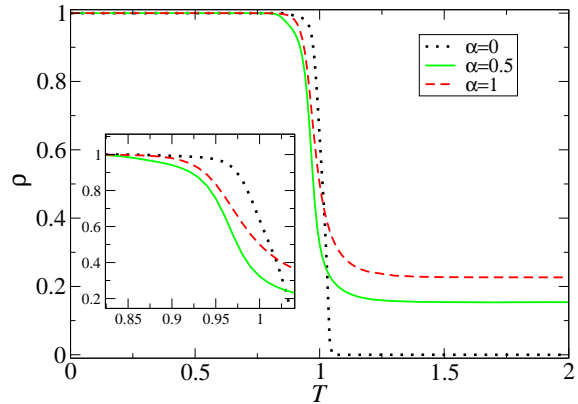}
\caption{Asymptotic cooperation level ($\rho$) versus temptation to defect ($T$) in a square lattice for the weak Prisoner's Dilemma. Each line corresponds to a different fraction of innovators ($\alpha$). The inset shows the $T$ region where the heterogeneous population, $\alpha=0.5$, has a cooperation level lower than any pure population.}
\label{fig_Tphase}
\end{figure}

Fig.~\ref{fig_Tphase}, based on the weak Prisoner's Dilemma ($S=0$), shows the $\rho-T$ graph for the following cases: fully imitative ($\alpha=0$), fully innovative ($\alpha=1$), and equally mixed population ($\alpha=0.5$). The behavior in the structured population is different from the well-mixed case, especially for $T>0.9$.

It is worth to note that, in the square lattice, even if the behavior of the mixed population stays, usually, between the two pure cases (i.e. $\alpha=0$ and $\alpha=1$), some values of $T$ can lead to different scenarios. In particular, in the range $0.8<T<1.03$, detailed in the inset of Fig.~\ref{fig_Tphase}, the heterogeneous population exhibits the lowest cooperation value among the three presented models. Surprisingly, we find that in this region cooperation is higher when the population is composed of only one kind of agent (i.e. full imitation or full innovation). In addition, we note that the considered range of $T$ contains the critical point of the phase transition from cooperation to defection in the full imitative model \cite{Szabo2007}, suggesting that a heterogeneous population undergoes a faster transition than a homogeneous one. 

\begin{figure}
\includegraphics[width=7.5cm]{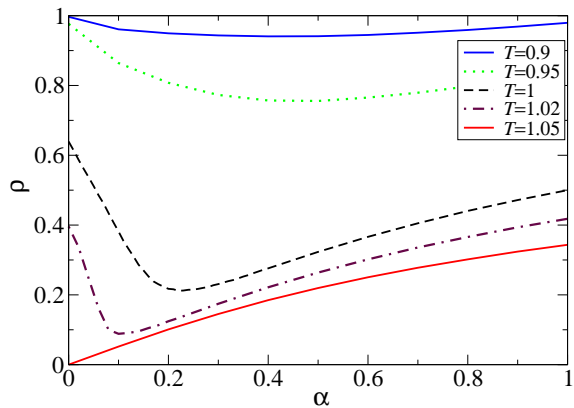}
\caption{ Asymptotic cooperation level, $\rho$, for increasing fraction of innovators ($\alpha$) in the square lattice. The behavior is quite different from the well-mixed population. In the region $T>1.04$ it is always better a fully innovative population. But for $0.8<T<1.04$ there is a always minimum value of $\rho$ for mixed populations.}
\label{fig_alpha}
\end{figure}

Fig.~\ref{fig_alpha} shows how the final cooperation level varies, as we increase the number of innovative agents for a given $T$ value ($S=0$). There are regions where the dependence with $\alpha$ is not trivial, especially around $T=1$ (i.e. near the edge between the Prisoner's Dilemma and the Stag-Hunt game). Notably, in this region, the mixing of different updating rules tends to reduce cooperation. This effect is especially strong near $T=1.04$, where the imitative model shows a phase transition \cite{Szabo2007}. We note that this drop in cooperation for mixed updating rules was also observed in \cite{Xu2017}, which considers a very different setting, the Public Goods Game with a different kind of innovative SRP (win-stay-lose-shift), mixed with the imitation model. Also, \cite{Danku2018} showed that mixed strategies coupled with co-evolutionary processes can lead to spontaneous cyclic dominance and diverse complex patterns in the population.

\begin{figure}
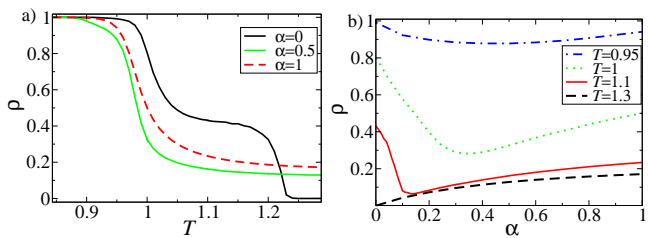

\includegraphics[width=4.2cm]{5a.eps}
\includegraphics[width=4.2cm]{5b.eps}
\caption{ Asymptotic cooperation level ($\rho$) in a triangular lattice. In a) we show the dependence in $T$ while in b) we show its dependence in $\alpha$. As we can see, the mixing of update rules can lead to a drop in cooperation for $0.95<T<1.25$. The behavior is qualitatively similar to the square lattice.}
\label{fig_Ttriang2}
\end{figure}

To further back our claims, we analyze the same setting in a triangular lattice with periodic boundary conditions. We see in Fig.~\ref{fig_Ttriang2}a) the same qualitative behavior observed in the square lattice, i.e., there is a drop in cooperation level near the phase transition point of the system when we mix updating rules. We see that for $0.95<T<1.25$, the mixing of strategies only decreases cooperation. The effect disappears after $T>1.25$, when cooperation is already extinct for the fully imitative population. Fig.~\ref{fig_Ttriang2}b) shows the behavior as we increase $\alpha$. As in the square lattice, $\rho$ reaches a minimum for small $\alpha$ values near the range of $T$ where the imitative model has a drop in cooperation.

\begin{figure}
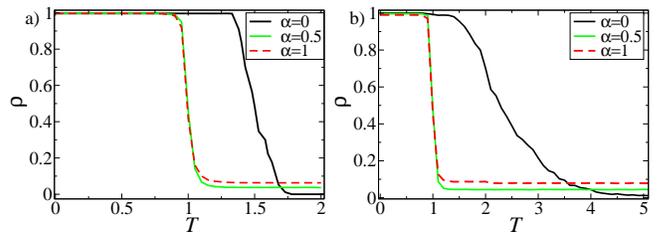

\includegraphics[width=4.2cm]{6a.eps}
\includegraphics[width=4.2cm]{6b.eps}
\caption{ Asymptotic cooperation level ($\rho$) as we increase $T$ for a random network in a) and a scale-free network in b).The drop in cooperation for the mixed population, compared to pure populations, happens in both topologies.}
\label{fig_Tcomplex}
\end{figure}

We also ran simulations of similar settings in a random and a scale-free network with average connectivity degree of 2.7, generated using the Krapivsky-Redner algorithm \cite{krapivsky_pre01}. Note that the scale-free network is a very famous case of spatial reciprocity when the imitation rule is used \cite{Szabo2007}, and at the same time it is know that the pure innovative rule destroys this reciprocity effect \cite{Amaral2016b}. The results are shown in Fig.~\ref{fig_Tcomplex}a) for the random and Fig.~\ref{fig_Tcomplex}b) for the scale-free network. The same qualitative effect was observe in these two topologies. As we approach a value o $T$ where cooperation drops for the pure imitative or innovative model, here $T=1$, cooperation from the mixed model drops below the value of any pure model. It is interesting to note the same effect in all these different topologies, as it points out to a general behavior.

\begin{figure}
\includegraphics[width=7.5cm]{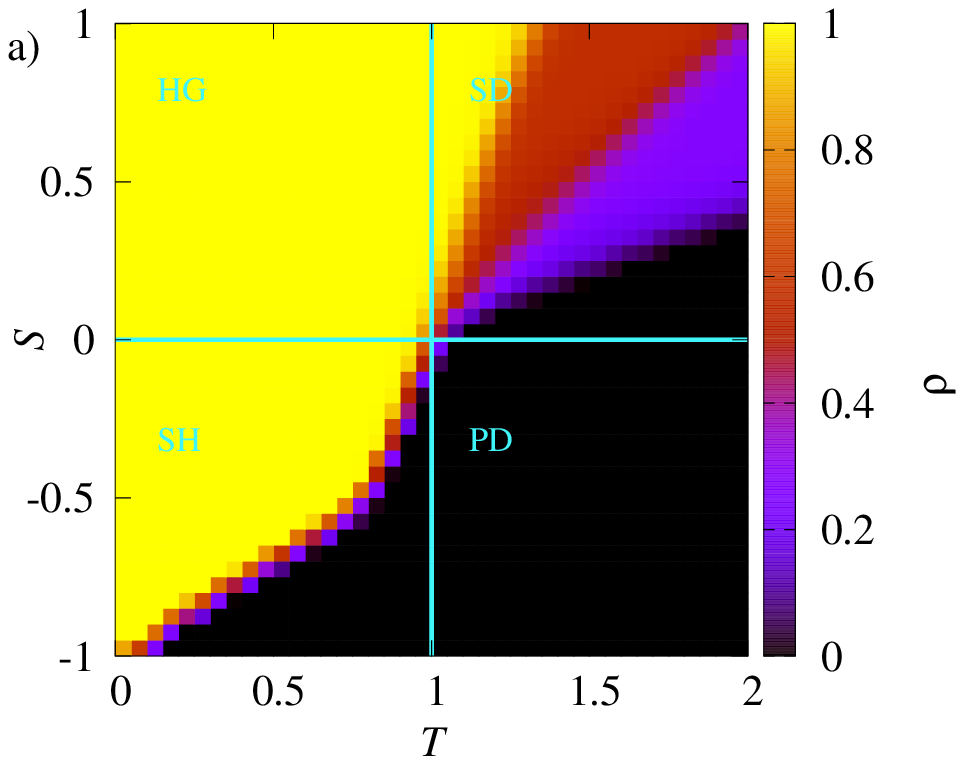}\vspace{0.5cm} 
\includegraphics[width=7.5cm]{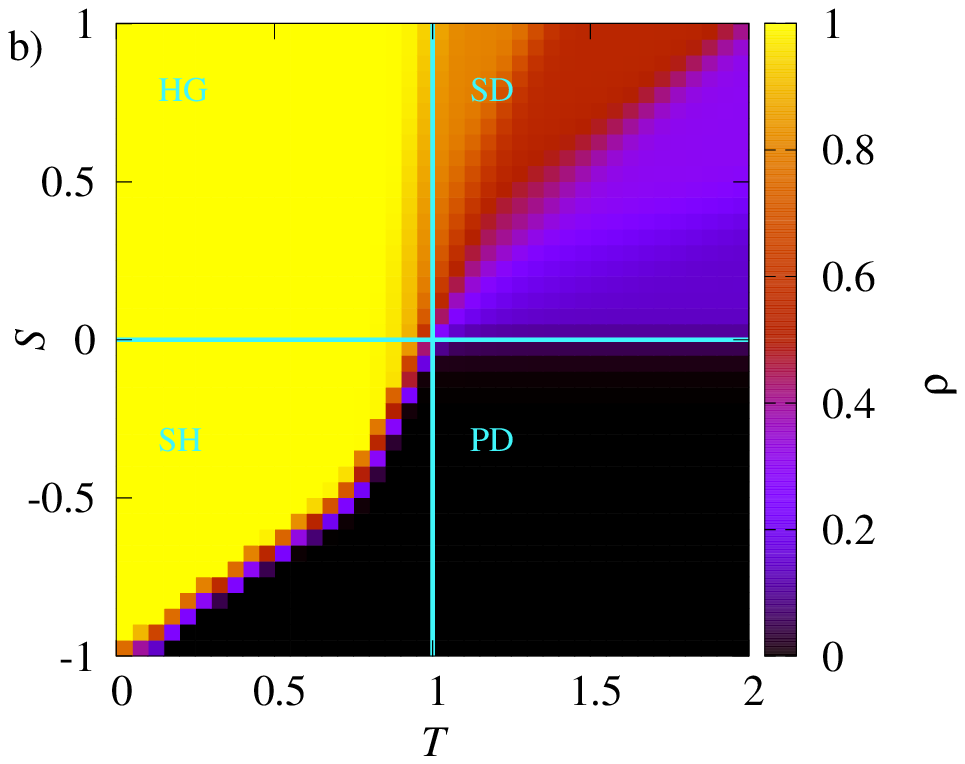}\vspace{0.5cm} 
\includegraphics[width=7.5cm]{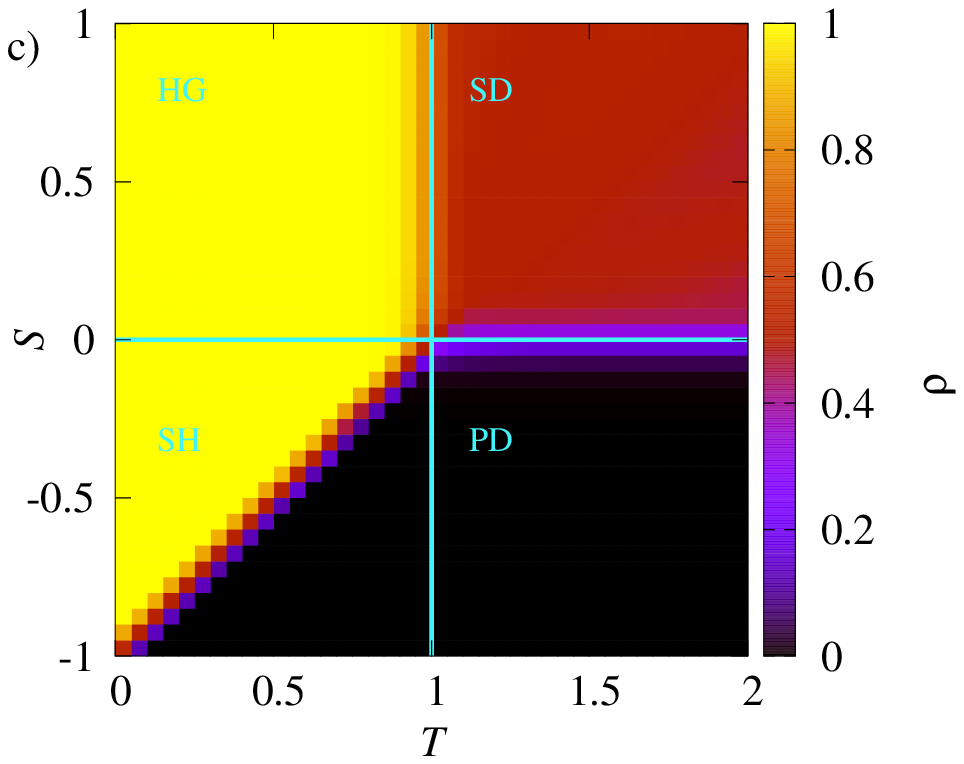} 
\caption{ Asymptotic cooperation level (colors) in the whole $T-S$ parameter space for the square lattice. We have a) pure imitation model ($\alpha=0$), b) the mixture of imitation and Logit dynamics ($\alpha=0.5$) and c) the pure Logit ($\alpha=1$) model. The main differences happens in the SH and SD regions.}
\label{fig_alphaphase}
\end{figure}

Next we present the asymptotic levels of cooperation in the full $T-S$ parameter space for the square lattice. The imitative model (i.e. $\alpha=0$) is presented in Fig.~\ref{fig_alphaphase}a), the heterogeneous population (i.e. $\alpha=0.5$) in Fig.~\ref{fig_alphaphase}b), and the fully innovative population (i.e. $\alpha=1$) in Fig.~\ref{fig_alphaphase}c). In this parametrization ($R=1$, $S=0$), each quadrant of the parameter space corresponds to one specific game: Harmony Game (HG), Snow-Drift (SD), Prisoner's Dilemma (PD) and Stag-Hunt (SH), in a clockwise fashion. Note that the pure cases differ mainly in the SH and SD regions, and the heterogeneous population leads to a behavior that is, usually, in between the two pure cases.

\begin{figure}
\includegraphics[width=7.5cm]{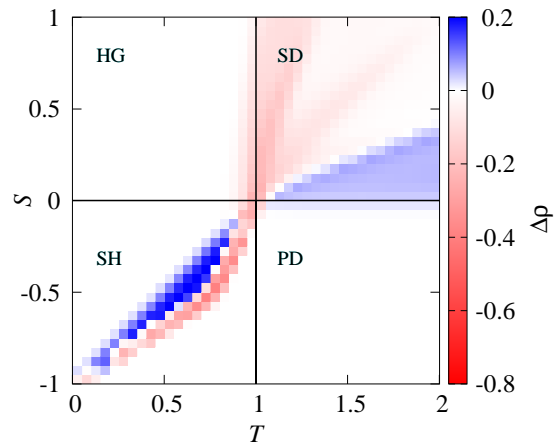} \hspace{.01ex}
\caption{Difference in asymptotic cooperation level (colors) between the mixed model ($\alpha=0.2$), and the average between the Logit and imitation model. The mixed model behaves differently from just the average of the two models, mainly in the SD and SH regions. }
\label{fig_deltaphase}
\end{figure}

With the aim to compare the mixed and pure cases in the square lattice, Fig.~\ref{fig_deltaphase} shows the difference in the final cooperation fraction between the heterogeneous model ($\rho_{mix}$), and the average value of the pure imitative ($\rho_{imt}$) and pure innovative ($\rho_{inv}$) models, i.e.
\begin{equation}
 \Delta \rho= \rho_{mix}-[(1-\alpha)\rho_{imt}+\alpha\rho_{inv}].
\end{equation}
This is particularly useful for observing whether there is any non-linear phenomenon. If innovative and imitative agents did not influence one another, it would be expected that $\Delta \rho=0$, as the mixing of the two would behave as just the average of the two pure models. Here we use $\alpha=0.2$, as this is the region where the mixed model differs most from the pure models. Note that the mixed model is mainly different from the average in the diagonal ($S=T-1$), with specific regions where the mixing can increase the cooperation in even $0.2$, or lower it in $-0.8$ for the SH region. On the other hand, there are smaller positive and negative differences through all the SD region. The Prisoner's Dilemma and Harmony Game regions are almost unchanged, except near the line $S=0$ (weak-Prisoner's Dilemma).

\begin{figure}
\includegraphics[width=7.5cm]{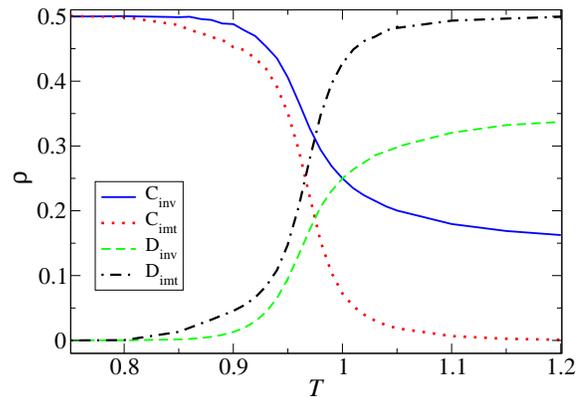}	
\caption{Asymptotic cooperation and defectors fraction for the two types of agents, innovative and imitative, in the square lattice. Here $S=0$ and $\alpha=0.5$. While innovative cooperators ($C_{inv}$) can survive for high $T$, conversely, it is the imitative defectors ($D_{imt}$) that fare better when $T>1$. Similar behavior occurs for all $\alpha$ values.}
\label{fig_indvpop}
\end{figure}

In order to understand how the mixing can be detrimental to cooperation, we study the subpopulation of innovative cooperators ($C_{inv}$) and imitative cooperators ($C_{imt}$) separately. Fig.~\ref{fig_indvpop} reports the four subpopulations (including innovative and imitative defectors) for the mixed case, $\alpha=0.5$. Note that imitative cooperators follow the usual behavior expected for a fully imitative population, i.e. they are almost extinguished for $T>1.04$, while innovative cooperators survive. But unexpectedly, while there are some innovative defectors, it is the imitative defectors that fare better for higher $T$ values. This result strongly suggests that the imitative behavior favors cooperation for $T<1$ and defection for $T>1$, while the innovative behavior has a smaller effect in this regard. We stress that such behavior is consistent for all values of $\alpha$.

\begin{figure}
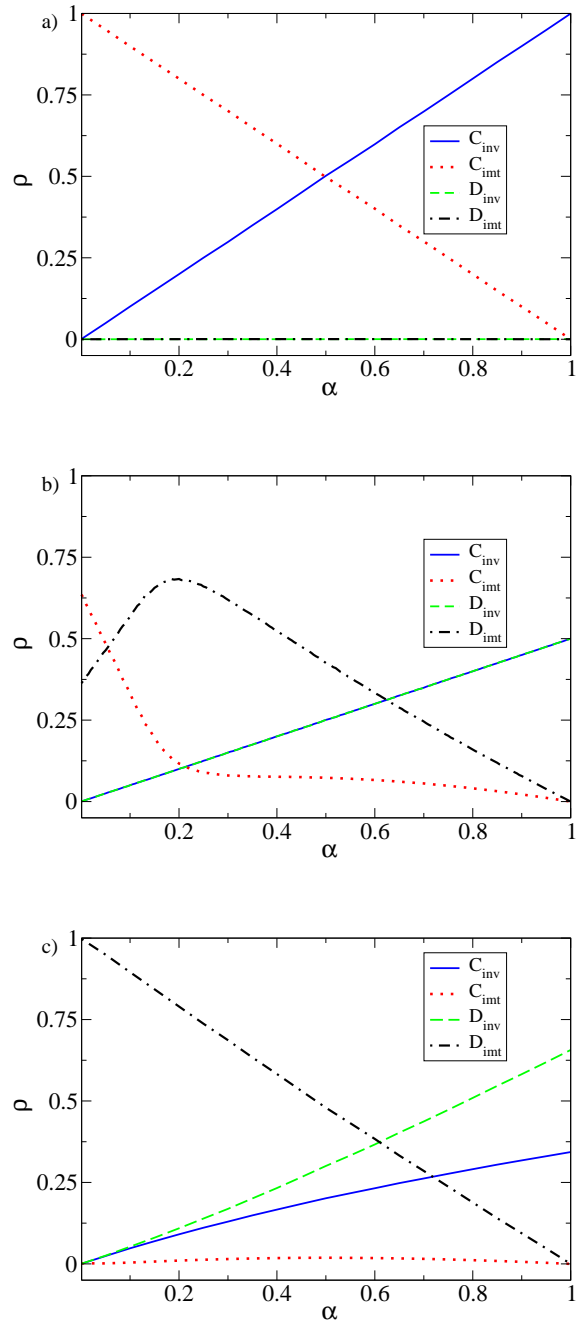

\includegraphics[width=7.5cm]{10a.eps}\vspace{0.9cm}
\includegraphics[width=7.5cm]{10b.eps}\vspace{0.9cm}
\includegraphics[width=7.5cm]{10c.eps}
\caption{Asymptotic fraction of the four types of agents (imitative and innovative cooperators and defectors) as $\alpha$ is increased for a) $T=0.8$, b) $T=1$ and c) $T=1.05$. The behavior is linear with $\alpha$ in a) and b). However, For $T=1$ there is a non-linear behavior, with an increase of $D_{imt}$ until $\alpha=0.2$.}
\label{fig_popalpha}
\end{figure}

In the same spirit, we present in Fig.~\ref{fig_popalpha} the four subpopulations as we continuously vary $\alpha$ for 3 different values of $T$. In Fig.~\ref{fig_popalpha}a), we have low temptation, $T=0.8$, where there is no defection and $C_{inv}$ grows, while $C_{imt}$ drops, linearly with $\alpha$. In Fig.~\ref{fig_popalpha}c), the temptation value is high (i.e. $T=1.05$) and then the same linear behavior occurs for defectors (although now there are some innovative cooperators that can survive for high $T$). The most interesting effect nevertheless occurs for intermediate values of $T$. Fig.~\ref{fig_popalpha}b), shows results for $T=1$, where a non-linear behavior emerges for imitative agents. We see that, as expected, the increasing in $\alpha$ (i.e. total fraction of innovative agents) is detrimental to imitative cooperators. However, remarkably, imitative defectors take profit from that, growing to a peak at $\alpha=0.2$. The mixing of updating rules favors defection for this range of $\alpha$ values (specifically imitative defectors). It is worth to emphasize that the described phenomenon is not intuitive. The increasing in innovative agents makes the subpopulation of imitative defectors sharply grow until $20\%$ of the lattice is composed of innovative agents.

\begin{figure}
\includegraphics[width=7.5cm]{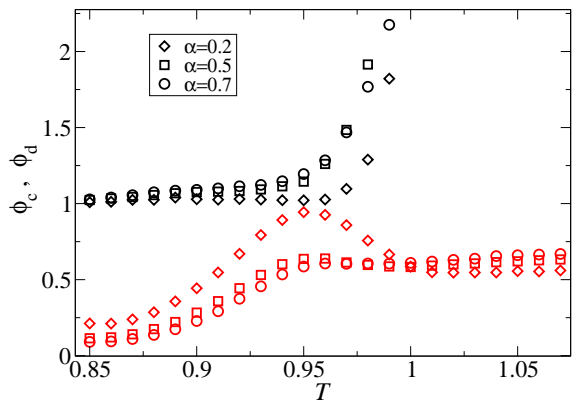} 
\caption{Ratio between the normalized fraction of innovative and imitative agents in the square lattice for three different values of $\alpha$. The ratio of cooperators is shown in black and defectors in red. Notice that $\phi_C$ grows with $T$, while $\phi_D$ is always smaller than $1$.}
\label{fig_ratio}
\end{figure}

Fig.~\ref{fig_ratio} reports the ratio between innovative and imitative agents of each strategy ($\phi_C$ and $\phi_D$ )in the region near $T=1$. To compare different fractions of innovative agents for different levels of $\alpha$, we normalize each population, dividing  the fraction of innovative cooperators by $\alpha$, and the fraction of imitative cooperators by $1-\alpha$ (and doing the same for defectors). This is done to prevent the oversampling of innovative agents in a scenario with high $\alpha$ values, i.e.
\begin{equation}
\phi_C = \dfrac{ C_{inv} } {\alpha} \dfrac{1-\alpha}{C_{imt}}~.
\end{equation}
In doing so, we can see that there is a general behavior in each population that is independent of $\alpha$. For $T<1$ the ratio $\phi_C$ is close to $1$, as cooperators from both types dominate the population. Although the total number of cooperators decreases as we increase $T$, the ratio between innovative and imitative cooperators keep increasing, indicating that innovative ones have the advantage. At the same time, $\phi_D$ varies for $1<T<1.04$ but is always below $1$ for the whole $T$ range. In other words, imitative sites will tend to be defectors, regardless of the total number of defectors. This general behavior occurs for any value of $\alpha$.

\begin{figure}
\includegraphics[width=4.2cm]{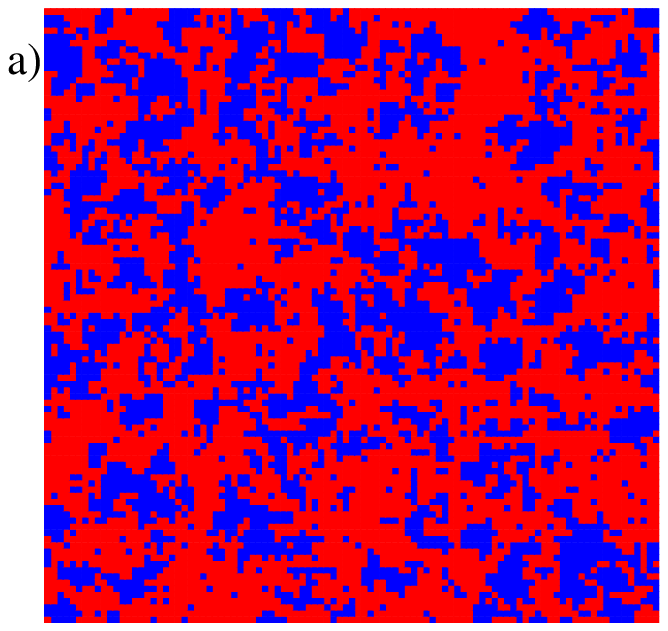}
\includegraphics[width=4.2cm]{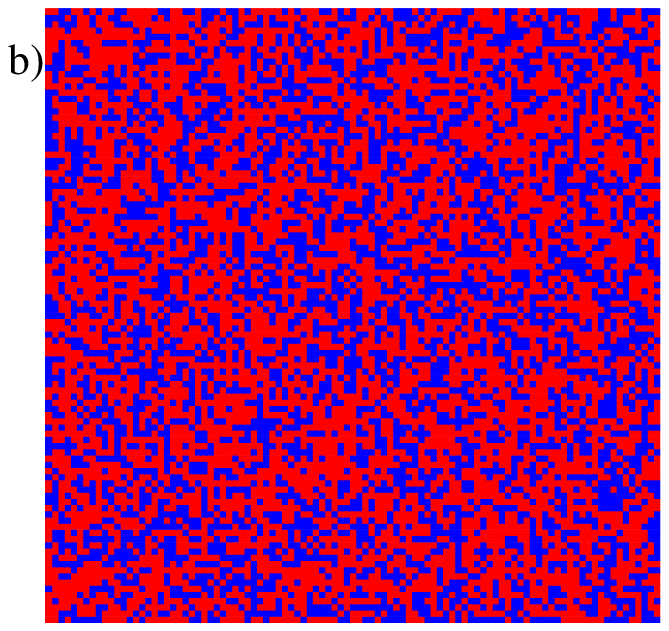}
\caption{Snapshots of the square lattice with $T=1$ for a) $\alpha=0.1$ and b) $\alpha=0.7$. As we mix innovators, the imitative cooperator islands gets dissolved, leading to a drop in total cooperation. Cooperators are shown in blue and defectors in red.}
\label{fig_snapsimple}
\end{figure}

Lastly, we analyze the snapshots of the square lattice to better understand this phenomenon on a microscopic level. Note that the Monte Carlo method is probabilistic, and accurate results are dependent on sufficiently large averages \cite{landau_00, binder_rpp97}. Nevertheless, looking at frames of the lattice, after the system has reached dynamical stability, can lead us to valuable insights. Those snapshots are shown in Fig.~\ref{fig_snapsimple}. We remind that the pure imitative model has a phase transition in $T=1.04$ \cite{Szabo2007}, near this region cooperation is mainly sustained because cooperators tend to form compact clusters to support each other \cite{Nowak1992a}. This behavior can be seen in Fig.~\ref{fig_snapsimple} a), where $\alpha=0.1$ and most of the population is imitative. At the same time, the pure innovative Logit model has a higher fraction of cooperators for $T=1.04$. However, in this case, cooperators do not form compact clusters. Instead, they spread out in the lattice and cooperation is sustained because of other mechanisms related to second order spatial effects as seen in \cite{Amaral2017, Amaral2016b}. Mixing both models, innovative cooperators spread trough the lattice and, in turn, imitative cooperators are not able to form clusters to protect themselves. At the same time, imitative defectors manage to invade cooperators from both sub-populations, leading to the downfall of cooperation. This can be seen in Fig.~\ref{fig_snapsimple}, as we increase $\alpha$ the clusters tend to dissolve. The process of dissolving the cooperator islands is gradual and continuous in $\alpha$. This is highly dependent on the parameters and just a small fraction of innovators can destroy the clusters when $T$ is near the phase transition point. The mixing of both updating rules near the critical point manages to neutralize the mechanism for maintaining cooperation from both imitative and innovative models. This is a robust mechanism, happening in the square and triangular lattice, as well as in random and scale-free networks. Nevertheless, it is important to keep in mind that this phenomenon happens for a specific range of parameters in the $T-S$ plane, near the phase transition of the pure imitative model.

\section{Conclusion}
\label{conclusion}
In this work, we investigate the evolutionary dynamics of heterogeneous populations, whose interactions are based on dilemma games.
In particular, our populations are composed of two kinds of agents, i.e. innovators and imitators. In principle, the main difference between them is related to the information source they use to modify their strategy, e.g. from cooperation to defection (or vice versa). Notably, innovators can estimate the potential gain they would receive when changing strategy, under the hypothesis that those of their neighbors remain constant. On the other hand, imitators take decisions by copying one randomly selected neighbor, depending on their payoff difference. 
As result, innovators are able to adopt even strategies that do not exist in their neighborhood, while imitators cannot do the same.

Innovation is an issue of paramount relevance in a number of systems, spanning from social to biological phenomena. Thus, it is expected to have an impact also in evolutionary games.
In order to shed further light on this aspect, the proposed model aims to analyze the influence of innovation by considering the updating mechanisms, i.e. the processes that allow agents to change strategy.
To this end, we first studied the dynamics of a population in the mean-field case, so that we were able to solve the model analytically.
Then, we performed numerical simulations considering agents arranged on a regular square and triangular lattice, as well as in random and scale-free networks.

The well-mixed case has a transition at $T=1$, where $\rho=0.5$, for any fraction of innovators. We found that, if $T<1$, imitation supports cooperation while, if $T>1$, innovation supports cooperation. This behavior was shown to be monotonous with the fraction of innovators only for the well-mixed case.
The structured case usually shows a behavior in between the pure kinds (full imitation and full innovation), although it is not a linear relation, i.e. $\langle\rho_{imt}+\rho_{inv}\rangle\neq \rho_{mix}$.
On the other hand, remarkably, we found that cooperation has a non trivial behavior for the heterogeneous population near phase transition points. For the square lattice, in the region $0.8<T<1.04$ there is always a minimum level of cooperation for any population mixing ($0<\alpha<1$). The triangular lattice, random and scale-free networks also show a similar behavior, i.e. cooperation drops near the phase transition point of each topology when we mix strategy updating rules.
Specifically, near the transition from cooperation to defection, homogeneous populations perform better than heterogeneous ones in supporting cooperation.
We also note that this kind of behavior has been reported in investigations based on a different scenario (i.e. using the Public Goods Games, mixing imitative and win-stay-lose-shift updates).

We obtained compelling evidences, that suggest this behavior is due to the interaction of innovative and imitative agents in heterogeneous populations.
In addition, lattice snapshots and the ratio of innovative/imitative agents indicate that near $T=1$, innovative cooperators destroy the spatial reciprocity while, at the same time, imitative defectors can invade both populations of cooperators. 
The mixing of two updating rules can destroy both mechanisms that sustain cooperation in each of the two pure cases.

The results of our investigations confirm that innovation plays a non-trivial role in evolutionary games. Diversity and heterogeneity  usually increase cooperation due to assortative effects. However, we have seen that this may not always be the case, as in some particular conditions mixed strategy revision rules can lead to lower cooperation. 
\begin{acknowledgments}
This research was supported by the Brazilian Research Agency CNPq, process 150524/2017-0. MAJ would like to acknowledge support by the H2020-645141 WiMUST project.
\end{acknowledgments}

\bibliographystyle{apsrev4-1}

\end{document}